\documentclass[preprint,12pt]{elsarticle}

\usepackage{amssymb}
\usepackage{dsfont}
\usepackage{threeparttable}
\newcommand{\I}{\mathrm{i}}
\newcommand{\G}{\mathbf{G}}
\newcommand{\R}{\mathbf{r}}
\newcommand{\K}{\mathbf{k}}

\journal{Journal of Sound and Vibration}

\begin{document}

\begin{frontmatter}

\title{In-plane vibrations of a rectangular plate: plane wave expansion modelling and experiment}

\author[label2]{A. Arreola-Lucas}
\address[label2]{Departamento de Ciencias B\'asicas, Universidad Aut\'onoma Metropolitana-Azcapotzalco, Av San Pablo 180, Col. Reynosa Tamaulipas, 02200 M\'exico DF, M\'exico.}
\ead{arreolaarturo@gmail.com}

\author[label1]{J. A. Franco-Villafa\~ne}
\address[label1]{Instituto de Ciencias F\'isicas, Universidad Nacional Aut\'onoma de M\'exico, A.P. 48-3, 62251 Cuernavaca, Morelos, M\'exico.}
\ead{jofravil@fis.unam.mx}

\author[label2]{G. B\'aez}
\ead{gbaez@correo.azc.uam.mx}

\author[label1]{R. A. M\'endez-S\'anchez\corref{cor1}}
\ead{mendez@fis.unam.mx}
\cortext[cor1]{Corresponding author, Tel. +52 55 562 27788; Fax: +52 55 562 27775.}

\begin{abstract}
Theoretical and experimental results for in-plane vibrations of a uniform rectangular plate with free boundary conditions are obtained. The experimental setup uses electromagnetic-acoustic transducers and a vector network analyzer. The theoretical calculations were obtained using the plane wave expansion method applied to the in-plane thin plate vibration theory. The agreement between theory and experiment is excellent for the lower 95 modes covering a very wide frequency range from DC to 20 kHz. Some measured normal-mode wave amplitudes were compared with the theoretical predictions; very good agreement was observed. The excellent agreement of the classical theory of in-plane vibrations confirms its reliability up to very high frequencies
\end{abstract}


\end{frontmatter}

\section{Introduction}
\label{Intro}

There is an increasing interest in the in-plane vibrations of plates. This is due to the fact that, in certain specialized applications, high-frequency vibrations appear. This commonly happens in data storage systems, in the kHz range, in which in-plane vibrations cause a problem in following narrow data tracks~\cite{HydeChangBaccaWickert}. These vibrations are also important in ship hull design since 
there is evidence that in-plane vibrations and high-frequency noise are strongly related~\cite{Gorman2004}. The in-plane modes also play an important role in the transmission of high frequency vibrations through a built-up structure~\cite{DuLiJinYangLiu}. Furthermore, in-plane modes can be used for non-destructive testing and evaluation of elastic constants~\cite{Larsson}. Finally, as in-plane vibrations appear at higher frequencies than transverse vibrations, finite element calculations are more difficult for the former. All these, and other problems not listed, have led to 
a renewed interest in the phenomenon of in-plane vibration of rectangular plates 
that cover several orders of magnitude from nanosystems to macrostrutures. 

There are several recent significant theoretical and numerical contributions to the study of the in-plane vibrations of plates~\cite{HydeChangBaccaWickert,Gorman2004,DuLiJinYangLiu,Gorman2004a,Graff,BardellLangleyDunsdon,LiuXing,Dozio,Gorman2005,Gorman,HuangMa}. However, experimental results have been, until recently, very scarce~\cite{Larsson,SchaadtSimonEllegaard}. There are two likely reasons of this fact: first, in-plane vibrations appear at high frequencies and second, the measurement of transverse vibrations is easier than the excitation and detection of in-plane modes~\cite{SchaadtSimonEllegaard,NievesGasconBayon,MaLin,JSV2010}. Such state of affairs started to change when electronic speckle pattern interferometry ---ESPI or TV holography--- made possible the measurement of in-plane modes~\cite{NievesGasconBayon,XuZhu,MonchalinAusselHeonJenBoundreaultBernier}. Thus, there is the need for research on the in-plane vibrations of plates to consolidate the classical theory of in-plane vibrations, 
especially at high frequencies, to contrast such theory with experimental results.

This work adds to the literature on the subject of in-plane vibration of plates which can be found in Ref.~\cite{Gorman}. 
In the next section, the plane wave expansion method, applied to the classical theory of in-plane vibrations of thin plates, is introduced. This method will be used to calculate the normal modes of a rectangular plate with two sets of boundary conditions, namely all its boundaries free (F-F-F-F) and all edges clamped (C-C-C-C). In Section~\ref{Sec:Experiment}, the experimental methodology to measure the in-plane vibrations of a rectangular plate with free boundaries, using electromagnetic-acoustic transducers and a vector network analyzer, is presented. In Section~\ref{Sec:Results} the theoretical normal-mode frequencies and wave amplitudes are compared with the experimental results, showing a very good agreement. Finally, in Section~\ref{Sec:Conclusions}, some conclusions are given.

\section{The plane wave expansion method for the in-plane wave equation}
\label{Sec:PWE}

\begin{figure}
\centering
\includegraphics[width=\columnwidth]{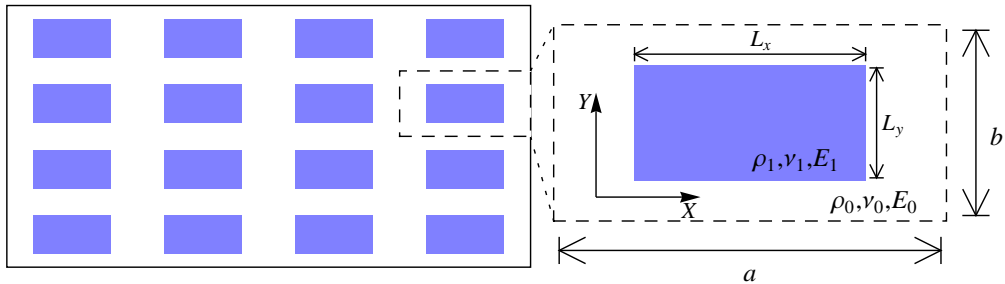}
\caption{Plane wave expansion method: a rectangular cell of sides $a$ and $b$ is repeated periodically on the plane. Each cell is composed of the plate of interest, in the center of the cell, surrounded by a host material. The elastic properties of the plate of interest have the subindex 1 while the elastic constants of the host material have the subindex 0. The host material mimics the vacuum, in a certain limit, that yields the free--end boundary conditions for the inner plate.}
\label{Fig:PWE_Scheme}
\end{figure}

The plane wave expansion method (PWE) refers to a computational technique to solve partial differential equations as an eigenvalue problem~\cite{JMP1974}. This method is popular among the photonic (phononic) crystal community to obtain the dispersion relation of artificial crystals~\cite{PLA2004,PLA2008,JELAST2013,Manzanares-MartinezRamos-MendietaBaltazar}. In a previous work~\cite{JSV2010} it was shown that the PWE method can be implemented to solve the out-of-plane Kirchhoff-Love equation for finite systems. As shown below, this numerical method is also useful to solve the in-plane wave equation for finite systems. The main difference between the plane wave expansion method and other numerical methods is that the boundary conditions are not imposed but are rather simulated by introducing a second medium with certain physical properties. A rectangular cell (see Fig.~\ref{Fig:PWE_Scheme}) of dimensions $a\times b$ will be used. The plate is located at the center of the cell surrounded by a host material 
that, for a plate with free ends, mimics the vacuum and, for a plate with clamped ends, mimics an extremely hard medium~\cite{JSV2010}. The unit cell is repeated periodically in both directions and its mechanical parameters are replaced by a Fourier series truncated at $N$ plane waves. In what follows, the PWE method as used to calculate the in-plane normal modes of plates with free-ends, will be described in detail.

The equations that govern the in-plane motion, in the classical theory of in-plane waves of a thin plate, are~\cite{Yu}
\begin{equation}\label{Equ:inplanewaves}
 \begin{array}{l}
  \displaystyle \frac{\partial N_{x}}{\partial x}+\frac{\partial N_{xy}}{\partial y}=\rho h\frac{\partial^2 u}{\partial t^2},\\
  \displaystyle \frac{\partial N_{xy}}{\partial x}+\frac{\partial N_{y}}{\partial y}=\rho h\frac{\partial^2 v}{\partial t^2},
 \end{array}
\end{equation}
where $h$ and $\rho$ are the thickness and density of the plate, respectively. The variables $u(x,y)$ and $v(x,y)$ are the displacements in the $X$ and $Y$ directions, respectively, while the plate stresses are
\begin{equation}\label{Equ:platestress}
 \begin{array}{l}
  \displaystyle N_{x}=C(e_{xx}+\nu e_{yy}),\\
  \displaystyle N_{y}=C(e_{yy}+\nu e_{xx}),\\
  \displaystyle N_{xy}=C(1-\nu)e_{xy},
 \end{array}
\end{equation}
where $\nu$ is Poisson's ratio and $C$ is the extensional rigidity given  by
\begin{equation}\label{Equ:extenrigidity}
 C=\frac{Eh}{1-\nu^2},
\end{equation}
with $E$ standing for Young's modulus. The strain-displacement relations are
\begin{equation}\label{Equ:straindisplacement}
 e_{xx}=\frac{\partial u}{\partial x},\quad e_{yy}=\frac{\partial v}{\partial y}\quad\mathrm{and}\quad e_{xy}=\frac{1}{2}\left(\frac{\partial u}{\partial y}+\frac{\partial v}{\partial x}\right).
\end{equation}

Since the mechanical properties of the system of Fig.~\ref{Fig:PWE_Scheme} are periodic, one can assume the following Fourier expansions for the coefficients that appear in Eqs.~(\ref{Equ:inplanewaves}) and~(\ref{Equ:platestress}):
\begin{eqnarray}
 C&=&\sum_{\G}\alpha_{\G}\exp(\I\G\cdot\R),\label{Equ:Cexp}\\
 C\nu&=&\sum_{\G}\beta_{\G}\exp(\I\G\cdot\R),\label{Equ:Cnuexp}\\
 \rho h&=&\sum_{\G}\eta_{\G}\exp(\I\G\cdot\R).\label{Equ:rhohexp}
\end{eqnarray}
Here $\R=(x,y)$ is the position and $\G=(G_x,G_y)=2\pi(p/a,q/b)$ is a vector of the reciprocal lattice with $p$ and $q$ integers. The displacements $u$ and $v$ are also periodic and can be written in terms of Fourier series as 
\begin{eqnarray}
 u&=&\exp(\I\K\cdot\R-\I\omega t)\sum_{\G}\phi_{\G}\exp(\I\G\cdot\R),\label{Equ:Uexp}\\
 v&=&\exp(\I\K\cdot\R-\I\omega t)\sum_{\G}\psi_{\G}\exp(\I\G\cdot\R),\label{Equ:Vexp}
\end{eqnarray}
where $\K=(k_x,k_y)$ is the wave vector and $\omega$ the angular frequency. 

Inserting the expansions~(\ref{Equ:Cexp}) to~(\ref{Equ:Vexp}) in Eqs.~(\ref{Equ:inplanewaves}) one gets, for each $\G$,
\begin{eqnarray}
 \sum_{\G'}\mathds{M}_{xx}(\G,\G')\phi_{\G'}+\mathds{M}_{xy}(\G,\G')\psi_{\G'}=\omega^2\sum_{\G'}\eta_{\G-\G'}\phi_{\G'},\label{Equ:inplane_x}\\
 \sum_{\G'}\mathds{M}_{yx}(\G,\G')\phi_{\G'}+\mathds{M}_{yy}(\G,\G')\psi_{\G'}=\omega^2\sum_{\G'}\eta_{\G-\G'}\psi_{\G'},\label{Equ:inplane_y}
\end{eqnarray}
where
\begin{equation}\label{Equ:Mxx_Def}
 \mathds{M}_{xx}(\G,\G')=\frac{1}{2}(\alpha_{\G-\G'}-\beta_{\G-\G'})(k_y+G'_y)(k_y+G_y)+\alpha_{\G-\G'}(k_x+G'_x)(k_x+G_x),
\end{equation}
\begin{equation}\label{Equ:Mxy_Def}
 \mathds{M}_{xy}(\G,\G')=\frac{1}{2}(\alpha_{\G-\G'}-\beta_{\G-\G'})(k_x+G'_x)(k_y+G_y)+\beta_{\G-\G'}(k_x+G_x)(k_y+G'_y),
\end{equation}
\begin{equation}\label{Equ:Myx_Def}
 \mathds{M}_{yx}(\G,\G')=\frac{1}{2}(\alpha_{\G-\G'}-\beta_{\G-\G'})(k_x+G_x)(k_y+G'_y)+\beta_{\G-\G'}(k_x+G'_x)(k_y+G_y),
\end{equation}
and
\begin{equation}\label{Equ:Myy_Def}
 \mathds{M}_{yy}(\G,\G')=\frac{1}{2}(\alpha_{\G-\G'}-\beta_{\G-\G'})(k_x+G'_x)(k_x+G_x)+\alpha_{\G-\G'}(k_y+G'_y)(k_y+G_y).
\end{equation}
Eqs.~(\ref{Equ:inplane_x}) and~(\ref{Equ:inplane_y}) can be written in matrix form as a generalized eigenvalue equation with an infinite number of columns and rows
\begin{equation}\label{Eq:inplane_Matriz}
\left( 
 \begin{array}{cc}
   \mathds{M}_{xx} &  \mathds{M}_{xy}\\
   \mathds{M}_{yx} &  \mathds{M}_{yy}
 \end{array}
\right)\left( 
 \begin{array}{c}
   \phi\\
   \psi
 \end{array}
\right)=\omega^2
\left( 
 \begin{array}{cc}
   \mathds{N} &  0\\
   0 &  \mathds{N}
 \end{array}
\right)\left( 
 \begin{array}{c}
   \phi\\
   \psi
 \end{array}
\right),
\end{equation}
with $\phi$ and $\psi$ vectors with entries $\phi_{\G'}$ and $\psi_{\G'}$, respectively, and
\begin{equation}\label{Eq:etaMatrix}
 \mathds{N(\G,\G')}=\eta_{\G-\G'}.
\end{equation}

\begin{figure}
\centering
\includegraphics[width=\columnwidth]{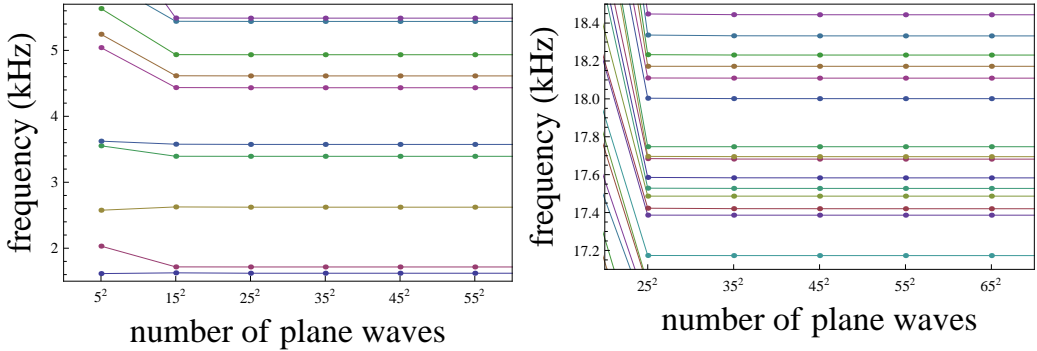}
\caption{Convergence of the plane wave expansion method for the lower normal modes (left panel) and for higher frequencies (right panel). The results correspond to a plate of $1474$~mm~$\times$~$355$~mm.}
\label{Fig:ConverPWE}
\end{figure}

The Fourier coefficients of Eqs.~(\ref{Equ:Cexp}) to~(\ref{Equ:rhohexp}) can be calculated analytically for regular shaped plates, such as the plate of Fig.~\ref{Fig:PWE_Scheme}. Plates with irregular shape can also be calculated but in this case the Fourier coefficients should be obtained numerically. The free or clamped boundary conditions can be obtained by defining a host material with $\rho_0\rightarrow 0$ or $\rho_0\rightarrow\infty$~\cite{JSV2010}, respectively. Although the generalized eigenvalue equation~(\ref{Eq:inplane_Matriz}) includes an infinite number of terms, in the numerical calculations the series were truncated to large partial sums. In  Fig.~\ref{Fig:ConverPWE} some eigenvalues for a rectangular plate with free-ends, as a function of the  of the number of plane waves, are plotted. It can be observed that $15\times 15$ plane waves are enough to get stable results at low frequencies but  $25\times 25$ plane waves are needed to obtain similar results at high frequencies.

In Tables~\ref{tab:Comp_theo_theo_FullF} and~\ref{tab:Comp_theo_theo_FullC} the dimensionless normal-mode frequencies of in-plane vibrations of a plate of different aspect ratios with free (F-F-F-F) and clamped (C-C-C-C) ends, respectively, are shown. 
The results of the PWE method show a very good agreement with the results available in the literature~\cite{Gorman2004,DuLiJinYangLiu,Gorman2004a,BardellLangleyDunsdon} with a difference less than 1.5\%. This shows that the PWE method can be used to calculate the normal-mode vibrations of in-plane waves.

\begin{table}
\caption{Dimensionless normal-mode frequencies $\Omega=\omega L_x\sqrt{\rho(1-\nu^2)/E}$ for the in-plane vibrations of a rectangular plate with different aspect ratios ($L_x/L_y$) and completely free (F-F-F-F) boundary conditions.}
\label{tab:Comp_theo_theo_FullF}
\centering
\resizebox{\columnwidth}{!}{
\begin{threeparttable}
\begin{tabular}{c*{8}{c}*{8}{c}}
 \hline
  && \multicolumn{7}{c}{aspect ratio $L_x/L_y=1$} && \multicolumn{7}{c}{aspect ratio $L_x/L_y=2$} \\
	\cline{3-9}\cline{11-17}
  && PWE && \multicolumn{2}{c}{Gorman I} && \multicolumn{2}{c}{Du and Bardell} && PWE && \multicolumn{2}{c}{Gorman I} && \multicolumn{2}{c}{Du and Bardell}\\
	\cline{3-3}\cline{5-6}\cline{8-9}\cline{11-11}\cline{13-14}\cline{16-17}
		Mode && $\Omega$ && $\Omega$ & \%Diff. && $\Omega$ & \%Diff. && $\Omega$ && $\Omega$ & \%Diff. && $\Omega$ & \%Diff.\\
	\hline
\phantom{2}1 && 2.321 && 2.320 & \phantom{2}0.043 && 2.321 & \phantom{2}0.000 && 1.958 && 1.956 & \phantom{2}0.112 && 1.954 & \phantom{2}0.205 \\ 

\phantom{2}2 && 2.474 && 2.472 & \phantom{2}0.081 && 2.472 & \phantom{2}0.081 && 2.963 && 2.960 & \phantom{2}0.101 && 2.961 & \phantom{2}0.068 \\ 

\phantom{2}3 && 2.474 && 2.472 & \phantom{2}0.081 && 2.472 & \phantom{2}0.081 && 3.270 && 3.268 & \phantom{2}0.061 && 3.267 & \phantom{2}0.092 \\ 

\phantom{2}4 && 2.631 && 2.628 & \phantom{2}0.114 && 2.628 & \phantom{2}0.114 && 4.728 && 4.726 & \phantom{2}0.042 && 4.726 & \phantom{2}0.042 \\ 

\phantom{2}5 && 2.990 && 2.988 & \phantom{2}0.067 && 2.987 & \phantom{2}0.100 && 4.786 && 4.784 & \phantom{2}0.042 && 4.784 & \phantom{2}0.042 \\ 

\phantom{2}6 && 3.454 && 3.452 & \phantom{2}0.058 && 3.452 & \phantom{2}0.058 && 5.208 && 5.208 & \phantom{2}0.000 && 5.205 & \phantom{2}0.058 \\ 

\phantom{2}7 && 3.724 && 3.724 & \phantom{2}0.000 &&& && 5.259 && 5.258 & \phantom{2}0.019 &&& \\ 

\phantom{2}8 && 3.724 && 3.724 & \phantom{2}0.000 &&& && 5.369 && 5.370 & -0.019 &&& \\ 

\phantom{2}9 && 4.305 && 4.306 & -0.023 &&& && 6.150 && 6.148 & \phantom{2}0.033 &&& \\ 

10 && 4.970 && 4.970 & \phantom{2}0.000 &&& && 6.448 &&  &  &&& \\ 

11 && 4.970 && 4.970 & \phantom{2}0.000 &&& && 6.597 && 6.596 & \phantom{2}0.015 &&& \\ 

12 && 5.047 && 5.046 & \phantom{2}0.020 &&& && 6.750 &&  &  &&& \\ 

13 && 5.257 && 5.258 & -0.019 &&& && 6.858 && 6.856 & \phantom{2}0.029 &&& \\ 

14 && 5.288 && 5.286 & \phantom{2}0.038 &&& && 7.452 &&  &  &&& \\ 

15 && 6.044 &&  &  &&& && 7.944 && 7.948 & -0.050 &&& \\
16 && 6.100 && 6.100 & 0.000 &&& && 8.546 &&  &  &&& \\
17 && 6.100 && 6.100 & 0.000 &&& && 8.667 &&  &  &&& \\ 
  \hline
\end{tabular}
\begin{tablenotes}[para,flushleft]
 {\large The results obtained with the plane wave expansion method (PWE) are compared with other methods. Gorman I, Du and Bardell results are taken from Refs.~\cite{Gorman2004},  ~\cite{DuLiJinYangLiu}~and~\cite{BardellLangleyDunsdon}, respectively.}
\end{tablenotes}
\end{threeparttable}
}
\end{table}

\begin{table}
\caption{Dimensionless normal-mode frequencies $\Omega=\omega L_x\sqrt{\rho(1-\nu^2)/E}$ for the in-plane vibrations of a rectangular plate with different aspect ratios ($L_x/L_y$) and completely clamped (C-C-C-C) boundary conditions.}
\label{tab:Comp_theo_theo_FullC}
\centering
\resizebox{\columnwidth}{!}{
\begin{threeparttable}
\begin{tabular}{c*{8}{c}*{8}{c}}
 \hline
  && \multicolumn{7}{c}{aspect ratio $L_x/L_y=1$} && \multicolumn{7}{c}{aspect ratio $L_x/L_y=2$} \\
	\cline{3-9}\cline{11-17}
  && PWE && \multicolumn{2}{c}{Gorman II} && \multicolumn{2}{c}{Du and Bardell} && PWE && \multicolumn{2}{c}{Gorman II} && \multicolumn{2}{c}{Du and Bardell}\\
	\cline{3-3}\cline{5-6}\cline{8-9}\cline{11-11}\cline{13-14}\cline{16-17}
		Mode && $\Omega$ && $\Omega$ & \%Diff. && $\Omega$ & \%Diff. && $\Omega$ && $\Omega$ & \%Diff. && $\Omega$ & \%Diff.\\
	\hline
\phantom{2}1 && 3.551 && 3.555 & -0.113 && 3.555 & -0.113 && 4.794 && 4.789 & \phantom{2}0.104 && 4.789 & \phantom{2}0.104 \\ 

\phantom{2}2 && 3.551 && 3.555 & -0.113 && 3.555 & -0.113 && 6.348 && 6.379 & -0.486 && 6.379 & -0.486 \\ 

\phantom{2}3 && 4.241 && 4.235 & \phantom{2}0.142 && 4.235 & \phantom{2}0.142 && 6.703 && 6.712 & -0.134 && 6.712 & -0.134 \\ 

\phantom{2}4 && 5.187 && 5.185 & \phantom{2}0.039 && 5.185 & \phantom{2}0.039 && 7.101 && 7.049 & \phantom{2}0.738 && 7.049 & \phantom{2}0.738 \\ 

\phantom{2}5 && 5.846 && 5.859 & -0.222 && 5.859 & -0.222 && 7.650 && 7.608 & \phantom{2}0.552 && 7.608 & \phantom{2}0.552 \\ 

\phantom{2}6 && 5.949 && 5.894 & \phantom{2}0.933 && 5.895 & \phantom{2}0.916 && 8.180 && 8.140 & \phantom{2}0.491 && 8.140 & \phantom{2}0.491 \\ 

\phantom{2}7 && 5.949 && 5.894 & \phantom{2}0.933 &&& && 8.987 && 8.998 & -0.122 &&& \\ 

\phantom{2}8 && 6.659 && 6.708 & -0.730 &&& && 9.535 && 9.515 & \phantom{2}0.210 &&& \\ 

\phantom{2}9 && 7.104 &&  &  &&& && 9.861 &&  &  &&& \\ 

10 && 7.104 && &  &&& && 10.640 &&  &  &&& \\ 

11 && 7.280 && 7.281 & -0.014 &&& && 11.260 && 11.250 & \phantom{2}0.089 &&& \\ 

12 && 7.572 && 7.597 & -0.329 &&& && 11.700 && 11.570 & \phantom{2}1.124 &&& \\ 

13 && 7.865 && 7.800 & \phantom{2}0.833 &&& && 11.900 && 11.740 & \phantom{2}1.363 &&& \\ 

14 && 8.610 &&  & &&& && 12.200 && & &&& \\ 

15 && 8.610 &&  & &&& && 12.430 && & &&& \\ 

16 && 8.698 && 8.718 & -0.229 &&& && 13.000 &&  & &&& \\ 

17 && 8.738 &&  &  &&& && 13.020 && 12.930 & \phantom{2}0.696 &&& \\ 
  \hline
\end{tabular}
\begin{tablenotes}[para,flushleft]
 {\large The results obtained with the plane wave expansion method (PWE) are compared with other methods. Gorman II, Du and Bardell results are taken from Refs.~\cite{Gorman2004a}, ~\cite{DuLiJinYangLiu}~and~\cite{BardellLangleyDunsdon}, respectively.}
\end{tablenotes}
\end{threeparttable}
}
\end{table}

\section{Experimental setup to measure in-plane waves}
\label{Sec:Experiment}

\begin{figure}
\centering
\includegraphics[width=0.8\columnwidth]{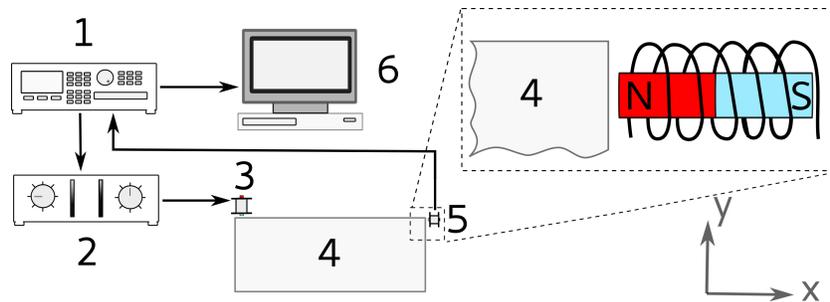}
\caption{Block diagram of the experimental setup. The vector network analyzer or VNA (1) sends a sinusoidal signal to the power amplifier (2). The amplified signal is sent to the EMAT exciter (3) which produces a mechanical vibration in the aluminum plate (4). The vibrations of the plate are measured by the EMAT detector (5). The measured signal by the EMAT is sent back to the VNA. The experimental data are finally collected in the computer (6) {\em via} diskette, serial port or GPIB.}
\label{Fig.Bloques}
\end{figure}

To measure the in-plane normal-mode frequencies for the rectangular plate with free-ends,  acoustic resonant spectroscopy (ARS)~\cite{EJP2012} is used. The experimental setup, shown in Fig.~\ref{Fig.Bloques}, consists of a vector network analyzer (VNA, Anritsu MS4630B); a high fidelity power amplifier (Cerwin-Vega CV900) and two electromagnetic-acoustic transducers (EMATs). The harmonic signal of frequency $f$, generated by the VNA, is sent to the power amplifier whose output is taken by the EMAT exciter. This transducer produces mechanical waves in the plate without mechanical contact. The EMAT detector measures the response of the plate in another location. The signal of this EMAT is registered directly by the VNA which yields the response of the plate at the frequency $f$. The frequency is changed to $f+\Delta f$, with $(\Delta f)/f \ll 1$. The response, as a function of the frequency, is then obtained and the peaks on it will correspond to the resonant frequencies of the plate.

The in-plane resonances of the plate are excited and detected selectively when using electromagnetic-acoustic transducers which consist of coils and magnets in particular configurations (see below). Since the transducers operate through eddy currents, they have to be located in the vicinity of a paramagnetic metal. Let us briefly recall the principles of operation of the EMATs; a detailed explanation of the operation of these transducers is given in Refs.~\cite{EJP2012,Emats}. The EMAT, as an exciter, operates in the following way:

\begin{enumerate}

\item{}A harmonic current, of frequency $f$, passes through the EMAT's coil; this generates a magnetic field $B(t)$ that varies in time  with the same frequency $f$.
 
\item{}Due to Faraday's law of induction, on any circuit of the paramagnetic material which is close to the EMAT's coil, local eddy currents are generated; these currents are also harmonic.

\item{}The eddy currents, induced on the paramagnetic metal, interact with the EMAT's magnet through the Lorentz force. 

\item{}The effect of the Lorenz force on the metal will become a mechanical vibration which travels along the plate. 

\end{enumerate}

The EMATs are also invertible, {\em i.e.} they can detect the vibrations of a paramagnetic metal. The electromagnetic-acoustic transducer, as a detector, operates in this way:

\begin{enumerate}

\item{}There will be a change of magnetic flux in the loops inside a vibrating paramagnetic metal when the EMAT's permanent magnet is near it.

\item{}The change of flux, by Faraday's law, will produce eddy currents in the metal; these currents oscillate with the frequency $f$ of the vibrating metal. 

\item{}The eddy currents will generate their own alternating magnetic field, which flows through the surface enclosed by the EMAT's coil. This variable flux will induce on the EMAT's coil an electromotive force measured by the VNA. 

\end{enumerate}

\section{Comparison between theory and experiment}
\label{Sec:Results}

In Fig.~\ref{Fig.Bloques} the configuration of the EMATs used to excite and detect the in-plane vibrations is shown. In this configuration the dipole moment axis of both the EMAT's coil and permanent magnet coincide and can be considered as the EMAT's axis; the axis of the exciter is parallel to the Y-axis while the axis of the detector is parallel to the X-axis. To decrease the possibility of missing resonances, {\em i.e.} to avoid measuring on a nodal line, the EMATs were located at the corners of the plate on the long side, where it is expected that the wave amplitude will have a maximum due to the free-ends. All the results presented in this section were obtained with this arrangement of EMATs. The main advantage of using the electromagnetic-acoustic transducers, in the configuration of Fig.~\ref{Fig.Bloques}, is that they are highly selective to in-plane vibrations. This result is very important since the measurement of in-plane vibrations is a difficult task. 

\begin{figure}
\includegraphics[width=\columnwidth]{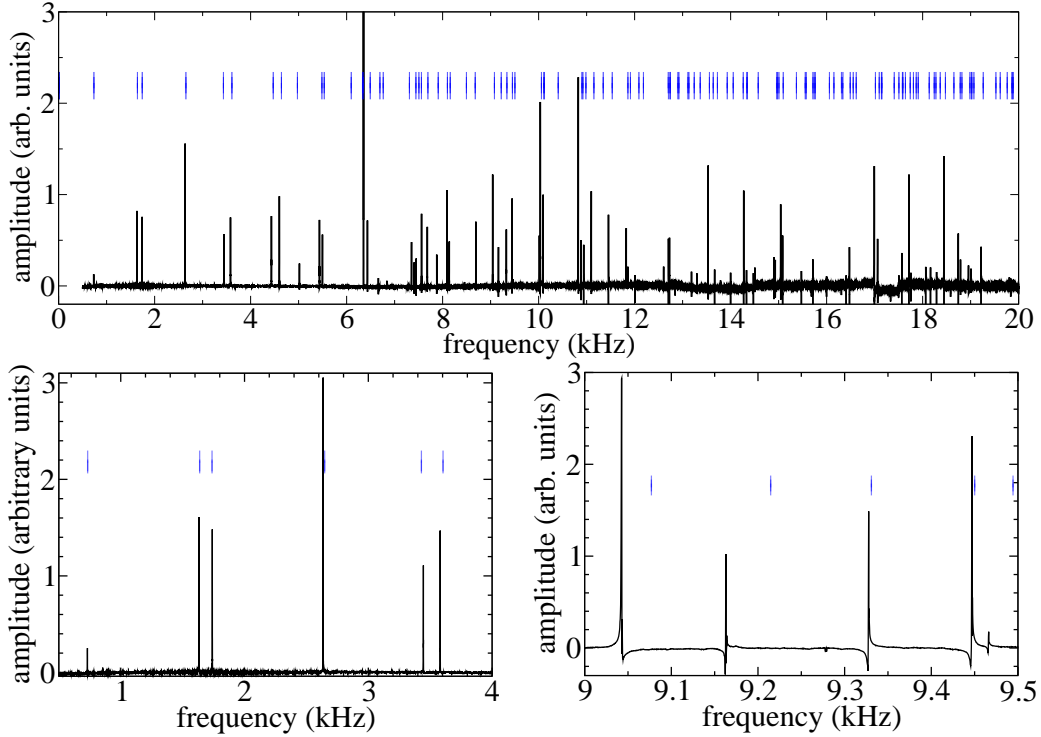}
\caption{Upper panel: DC to 20 kHz spectrum of a rectangular aluminum plate obtained with the setup of Fig.~\ref{Fig.Bloques}. The EMATs excite and detect selectively in-plane vibrations. The vertical lines correspond to the theoretical predictions obtained with the plane wave expansion method. In the lower panels amplifications, of the spectrum at low and intermediate frequencies, are given.}
\label{Fig:Comp_Theo_Exp}
\end{figure}

\begin{figure}
\centering
\includegraphics[width=0.8\columnwidth]{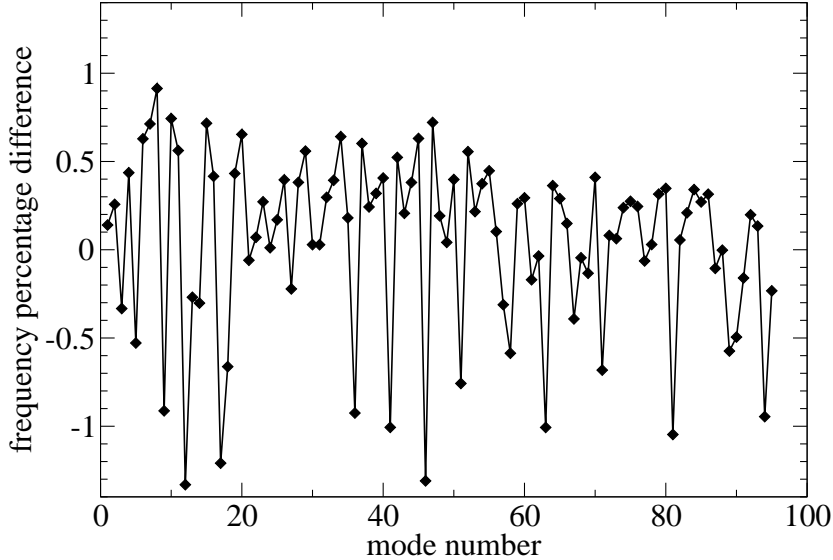}
\caption{Percentage difference between the experimental and theoretical normal-mode frequencies as a function of the normal mode number.}
\label{Fig:Error}
\end{figure}

\begin{figure}
\centering
\includegraphics[width=0.8\columnwidth]{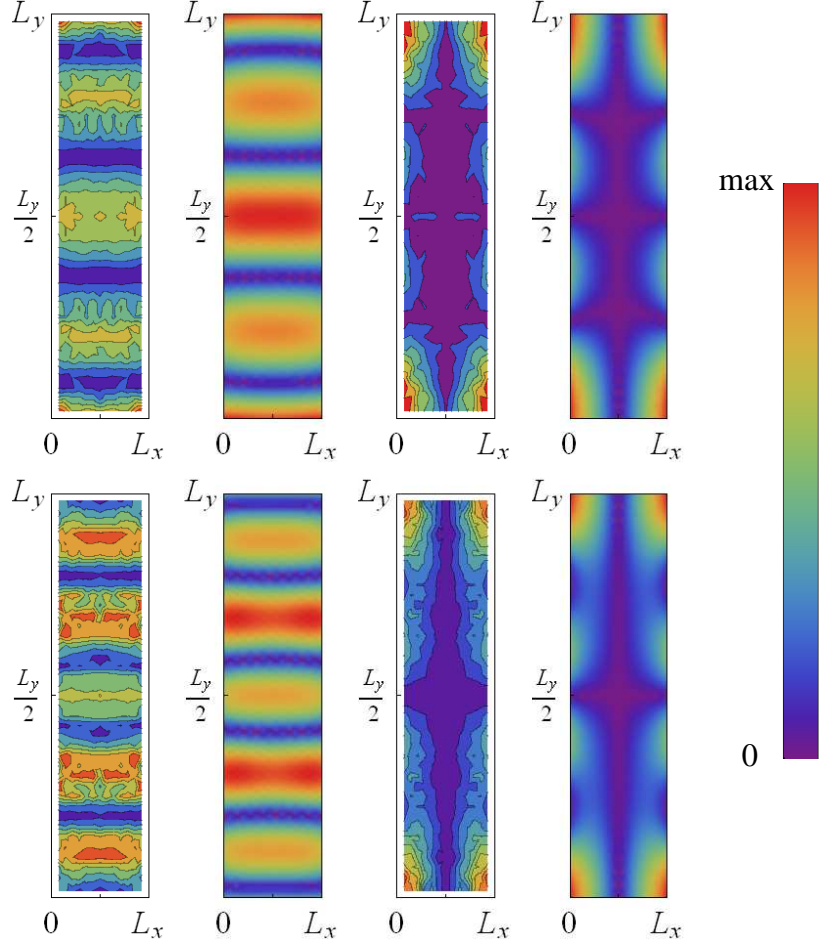} 
\caption{Theoretical and experimental wave amplitudes (absolute value) of the in-plane normal modes of the rectangular aluminum plate of $1474$~mm~$\times$~$355$~mm~$\times$~$6.35$~mm. The upper and lower parts correspond to modes 4 and 7 of Table~\ref{tab:Comp_theo_exp}, respectively. The first and second columns correspond to the EMAT measurement and PWE prediction of $u$, respectively; the third and fourth columns correspond to the EMAT measurement and PWE prediction of $v$, respectively. To avoid spurious border effects, the experimental wave amplitudes were not measured close to the boundary. On the right, the scale is shown.
}
\label{Fig:NormalModes}
\end{figure}

The spectrum of a rectangular aluminum plate of $1474$~mm~$\times$~$355$~mm and thickness $6.35$~mm, measured with the setup of Fig.~\ref{Fig.Bloques}, is given in Table~\ref{tab:Comp_theo_exp} and plotted in Fig.~\ref{Fig:Comp_Theo_Exp}. Something worthy to remark, as it can be seen in the lower panels of this figure, is that only in-plane resonances are observed. The theoretical prediction, obtained with the plane wave expansion method, using the best fit elastic constants of aluminum ($E = 71.1$~GPa and $\nu = 0.36$) and the measured density $\rho= 2708$~kg/m$^3$, is also given in Table~\ref{tab:Comp_theo_exp} and plotted in Fig.~\ref{Fig:Comp_Theo_Exp}. The agreement is excellent at low frequencies and, remarkably, also at high frequencies (see lower panels of this figure). The difference between the experiment and the theoretical predictions was also quantified and is plotted, as a function of the normal mode number, in Fig.~\ref{Fig:Error}. One can observe in this figure that the 
difference is always less than 1.4\%, a very reliable value for 
more than 90 normal modes. 

\begin{center}
\begin{table}
\caption{In-plane normal-mode frequencies of a rectangular aluminum plate of $1474$~mm~$\times$~$355$~mm~$\times$~$6.35$~mm with free ends (F-F-F-F) measured with acoustic resonant spectroscopy (ARS) and calculated with the plane wave expansion method (PWE).}
\label{tab:Comp_theo_exp}
\resizebox{\columnwidth}{!}{
\begin{tabular}{rrrrrrrrrrrr}
	\hline
	& & & & & & & & & & & \\
  Mode  & ARS & PWE & Diff. & Mode &  ARS & PWE & Diff. & Mode &  ARS & PWE & Diff. \\
        & (Hz) & (Hz)  & (\%)  &      & (Hz) & (Hz)  & (\%)  &      & (Hz) & (Hz)  & (\%)  \\ 
\hline
	& & & & & & & & & & & \\
 \phantom{2}1 &   731 &   730.0 & $-0.14$ & 33 & 10015 & 10054.4 & $ 0.39$ & 65 & 14910 & 14953.2 & $ 0.29$ \\ 
 \phantom{2}2 &  1632 &  1636.2 & $ 0.26$ & 34 & 10035 & 10099.3 & $ 0.64$ & 66 & 14938 & 14960.2 & $ 0.15$ \\ 
 \phantom{2}3 &  1740 &  1734.2 & $-0.33$ & 35 & 10094 & 10112.2 & $ 0.18$ & 67 & 15053 & 14994.0 & $-0.39$ \\ 
 \phantom{2}4 &  2634 &  2645.5 & $ 0.44$ & 36 & 10496 & 10398.9 & $-0.92$ & 68 & 15094 & 15087.1 & $-0.05$ \\ 
 \phantom{2}5 &  3445 &  3426.8 & $-0.53$ & 37 & 10827 & 10892.2 & $ 0.60$ & 69 & 15387 & 15366.5 & $-0.13$ \\ 
 \phantom{2}6 &  3580 &  3602.5 & $ 0.63$ & 38 & 10886 & 10912.4 & $ 0.24$ & 70 & 15485 & 15548.5 & $ 0.41$ \\ 
 \phantom{2}7 &  4434 &  4465.6 & $ 0.71$ & 39 & 10939 & 10973.9 & $ 0.32$ & 71 & 15674 & 15567.1 & $-0.68$ \\ 
 \phantom{2}8 &  4598 &  4640.0 & $ 0.91$ & 40 & 11098 & 11143.1 & $ 0.41$ & 72 & 15695 & 15707.7 & $ 0.08$ \\ 
 \phantom{2}9 &  5017 &  4971.2 & $-0.91$ & 41 & 11455 & 11339.7 & $-1.01$ & 73 & 15721 & 15730.9 & $ 0.06$ \\ 
           10 &  5435 &  5475.4 & $ 0.74$ & 42 & 11465 & 11525.0 & $ 0.52$ & 74 & 15726 & 15763.5 & $ 0.24$ \\ 
           11 &  5497 &  5527.9 & $ 0.56$ & 43 & 11825 & 11849.4 & $ 0.21$ & 75 & 16009 & 16052.9 & $ 0.27$ \\ 
           12 &  6167 &  6084.9 & $-1.33$ & 44 & 11859 & 11904.2 & $ 0.38$ & 76 & 16105 & 16144.5 & $ 0.25$ \\ 
           13 &  6350 &  6332.9 & $-0.27$ & 45 & 12005 & 12080.7 & $ 0.63$ & 77 & 16317 & 16306.6 & $-0.06$ \\ 
           14 &  6353 &  6333.8 & $-0.30$ & 46 & 12339 & 12177.4 & $-1.31$ & 78 & 16323 & 16327.8 & $ 0.03$ \\ 
           15 &  6435 &  6481.1 & $ 0.72$ & 47 & 12605 & 12695.9 & $ 0.72$ & 79 & 16429 & 16480.8 & $ 0.32$ \\ 
           16 &  6660 &  6687.7 & $ 0.42$ & 48 & 12705 & 12729.4 & $ 0.19$ & 80 & 16490 & 16547.3 & $ 0.35$ \\ 
           17 &  6839 &  6756.3 & $-1.21$ & 49 & 12732 & 12737.3 & $ 0.04$ & 81 & 16782 & 16606.3 & $-1.05$ \\            
           18 &  7352 &  7303.3 & $-0.66$ & 50 & 12844 & 12895.1 & $ 0.40$ & 82 & 17003 & 17012.4 & $ 0.06$ \\ 
           19 &  7405 &  7437.0 & $ 0.43$ & 51 & 13015 & 12916.4 & $-0.76$ & 83 & 17054 & 17089.7 & $ 0.21$ \\ 
           20 &  7449 &  7497.7 & $ 0.65$ & 52 & 13028 & 13100.4 & $ 0.56$ & 84 & 17078 & 17136.2 & $ 0.34$ \\ 
           21 &  7561 &  7556.5 & $-0.06$ & 53 & 13095 & 13123.3 & $ 0.22$ & 85 & 17100 & 17146.3 & $ 0.27$ \\           
           22 &  7681 &  7686.4 & $ 0.07$ & 54 & 13185 & 13234.3 & $ 0.37$ & 86 & 17348 & 17402.6 & $ 0.31$ \\ 
           23 &  7877 &  7898.4 & $ 0.27$ & 55 & 13300 & 13359.5 & $ 0.45$ & 87 & 17515 & 17496.5 & $-0.11$ \\ 
           24 &  8094 &  8094.9 & $ 0.01$ & 56 & 13541 & 13554.9 & $ 0.10$ & 88 & 17575 & 17574.6 & $-0.00$ \\ 
           25 &  8138 &  8151.8 & $ 0.17$ & 57 & 13675 & 13632.4 & $-0.31$ & 89 & 17682 & 17580.5 & $-0.57$ \\ 
           26 &  8457 &  8490.5 & $ 0.40$ & 58 & 13800 & 13719.1 & $-0.59$ & 90 & 17723 & 17635.3 & $-0.49$ \\ 
           27 &  8692 &  8672.7 & $-0.22$ & 59 & 13886 & 13922.2 & $ 0.26$ & 91 & 17763 & 17734.6 & $-0.16$ \\ 
           28 &  9042 &  9076.5 & $ 0.38$ & 60 & 14007 & 14048.2 & $ 0.29$ & 92 & 17766 & 17801.2 & $ 0.20$ \\ 
           29 &  9163 &  9214.2 & $ 0.56$ & 61 & 14281 & 14256.7 & $-0.17$ & 93 & 17832 & 17855.9 & $ 0.13$ \\ 
           30 &  9328 &  9330.7 & $ 0.03$ & 62 & 14339 & 14333.9 & $-0.04$ & 94 & 18070 & 17899.1 & $-0.95$ \\ 
           31 &  9447 &  9449.7 & $ 0.03$ & 63 & 14482 & 14336.2 & $-1.01$ & 95 & 18171 & 18128.7 & $-0.23$ \\ 
           32 &  9466 &  9494.1 & $ 0.30$ & 64 & 14515 & 14567.7 & $ 0.36$ &    &       &         &         \\               
\hline
\end{tabular}
}
\end{table}
\end{center}

The measurement of the components $u(x,y)$ and $v(x,y)$ of the in-plane normal-mode wave amplitudes is performed as follows. The in-plane vibrations of the aluminum plate are generated using an EMAT located at one end of the plate in the configuration of Fig.~\ref{Fig.Bloques}. This device is controlled by the VNA, which sends a signal sweeped around the resonant frequency of the wave amplitude to be measured. As it is well known, plotting the norm and phase of the response gives a circle in the complex plane; the radius of the circle is proportional to the wave amplitude. Measurements were carried out on a quarter of the plate in a rectangular grid. The EMAT measures the acceleration in the direction of the coil axis~\cite{JSV2012}. The wave amplitude is then measured using an EMAT located over the plate; the dipole axis of the magnet's EMAT is perpendicular to the X-Y plane while the coil axis is aligned along the X or Y direction to measure the deformation $u$ or $v$, respectively. To obtain the wave 
amplitude on the full plate, the measured data were reflected on the two axes of symmetry of the system. In Fig.~\ref{Fig:NormalModes} some normal-mode wave amplitudes are given and compared with those calculated with the plane wave expansion method; a very good agreement is obtained.

\section{Conclusions}
\label{Sec:Conclusions}

The in-plane vibrations of a rectangular aluminum plate, with free ends, have been studied both theoretically and experimentally. The experiments were performed using a vector network analyzer and electromagnetic-acoustic transducers. For the theoretical calculations, the plane wave expansion method, applied to the classical theory of in-plane waves, was used. The plane wave expansion method was first tested by comparing its results with other numerical methods with free and clamped boundary conditions. For 95 normal modes, within the range of DC-20~kHz, the agreement between theory and experiment is excellent since the difference is less than 1.4 percent. Some normal-mode wave amplitudes were measured and compared with the theoretical predictions; very good agreement was obtained. Thus, both the plane wave expansion method and the experimental setup based on electromagnetic-acoustic transducers could be very useful to study in-plane vibrations of thin plates with other shapes.

\section{Acknowledgments}

This work was partially supported by PAPIIT DGAPA-UNAM under project IN111311. AAL and JAFV were supported by a scholarship from CONACYT; AAL was partially supported by CONACYT under project 79613. Part of this work was written at the Universidad Polit\'ecnica de Valencia. We would like to thank Profs. A. L. Salas-Brito, J. S\'anchez-Dehesa and S. Ilanko for invaluable comments. 

\bibliographystyle{elsarticle-num}

\end{document}